\documentclass[prb,twocolumn,showpacs,preprintnumbers,amsmath,amssymb]{revtex4-1}

\usepackage{graphicx}
\usepackage{bm}

\newcommand{\be}{\begin{equation}}
\newcommand{\ee}{\end{equation}}

\def\bea{\begin{eqnarray}}
\def\eea{\end{eqnarray}}
\def\rmi{{\rm i}}
\def\rmexp{{\rm exp}}
\def\rmRe{{\rm Re}}
\def\rmIm{{\rm Im}}
\def\rmmin{{\rm min}}
\def\rmnem{{\rm nem}}
\def\rmCDW{{\rm CDW}}

\def\bra{\langle}
\def\ket{\rangle}

\begin{document}

\title{Competition between spin-induced charge instabilities in underdoped 
cuprates}
\author{Roland Zeyher$^a$}
\affiliation{$^a$Max-Planck-Institut f\"ur Festk\"orperforschung,
Heisenbergstrasse 1, D-70569 Stuttgart, Germany}

\author{Andr\'es Greco$^b$}
\affiliation{$^b$Facultad de Ciencias Exactas, Ingenier\'{\i}a y Agrimensura and
Instituto de F\'{\i}sica Rosario (UNR-CONICET),
Av. Pellegrini 250, 2000 Rosario, Argentina}

\date{\today}

\begin{abstract}

We study the static charge correlation function in an one-band 
model on a square lattice. The Hamiltonian consist of effective hoppings 
of the electrons between the lattice sites and the Heisenberg
Hamiltonian.
Approximating the irreducible charge correlation function by a single
bubble yields the ladder approximation for the charge correlation function. 
In this approximation
one finds in general three charge instabilities, two of them 
are due to nesting, the third one is the flux phase instability. 
Since these instabilities cannot explain the experiments in hole-doped
cuprates we have included in the irreducible charge correlation function
also Aslamasov-Larkin (AL) diagrams 
where charge fluctuations interact with products of spin fluctuations.
We then find at high temperatures a nematic or $d$-wave Pomeranchuk 
instability with a very small momentum. Its transition temperature decreases 
roughly 
linearly with doping in the underdoped region and vanishes near optimal
doping. Decreasing the temperature further a secondary axial 
charge-density wave (CDW) instability 
appears with mainly $d$-wave symmetry and a wave vector somewhat larger than
the distance between nearest neighbor hot spots. At still lower temperatures  
the diagonal flux phase instability emerges.  
A closer look shows that
the AL diagrams enhance mainly axial and not diagonal charge fluctuations
in our one-band model. This is the main reason why axial and not diagonal 
instabilities are the leading ones in agreement with experiment. 
The two instabilities due to nesting vanish already at very 
low temperatures and do not play any major role in the phase diagram.  
Remarkable is that the nematic and the axial CDW 
instabilities show a large reentrant behavior.
\end{abstract}


\maketitle

\section{introduction}

Cuprate superconductors show besides superconductivity and
antiferromagnetism phases with different modulations of the charge 
and spin density \cite{vojta09} including nematic \cite{fradkin10}
and liquid-crystal \cite{kivelson98} phases.
The striped or spin-charge ordered state in La-based compounds  
is well known. \cite{kivelson03}
More recently, x-ray scattering has shown the existence of incommensurate
CDW states which, different from stripes,
are not accompanied by spin order. These CDWs exist 
in several hole-doped cuprates, 
\cite{wu11, ghiringhelli12, chang12, achkar12, blackburn13, leboeuf13, 
blanco-canosa14, comin14, da-silva-neto14, 
hashimoto14, tabis14, gerber15, peng16, tabis17} and have many
properties in common: 
a) The modulation vector of the CDW is axial  
and decreases monotonically with increasing doping 
in contrast to the incommensurability of the spin 
fluctuations. \cite{yamada98,haug10,huecker11} 
b) In usual CDWs the electron and hole are localized  
at the same site, only their amplitude varies throughout the 
crystal. In contrast to that the electron and hole in the above cuprates are 
less localized and form a bound state with an internal 
$d$-wave symmetry. \cite{comin14,hamidian15} 
c) The cross-over temperature $T_{\rmCDW}$ to this new CDW state
lies well below the pseudogap temperature $T^*$ and shows a domelike 
shape  
\cite{blanco-canosa14} as a function of doping. Near its maximum 
a nematic transition line increases rapidly with decreasing doping. 
\cite{choiniere18}  

Theoretically it has been proposed that CDW instabilities, including 
nematic or $d$-wave Pomeranchuk instabilities, are caused either by 
the pseudogap phase \cite{atkinson15,comin14,atkinson16}  
and the related modification of the bandstructure or occur in the  
paramagnetic state due to antiferromagnetic exchange interactions. 
\cite{yamase00a,yamase00b,cappelluti99,metzner00,yamase07a,metlitski10a,husemann12,
bejas12,sachdev13,allais14,sau14,efetov13} 
In the second case mean-field calculations \cite{cappelluti99,bejas12,
sachdev13,allais14} 
yield a CDW instability to a flux state with a momentum near 
$(\pi,\pi)$ or an instability with a nesting vector connecting hot spots
along the diagonals. These instabilities are diagonal and not axial as in the
experiment. Calculations based on the spin fermion model found 
an instability with an axial wave vector connecting neighboring hot spots.
\cite{wang14}
Whether the employed simple ladder diagrams can produce a CDW with an axial 
momentum is presently not clear.  \cite{mishra15} A spin fermion model 
with overlapping hot spots may, however, yield an axial CDW instability. 
\cite{volkov16}  

A three-band Hubbard model treated in the mean-field approximation plus 
a charge interaction induced by product of spins (Aslamasov-Larkin or AL
diagrams) yielded a CDW with the correct $d$-wave symmetry
and a wave vector related to hot spots. \cite{yamakawa15,tsuchiizu16}
Whether this three band model can be 
applied to cuprates is unclear because the formation of Zhang-Rice singlets
\cite{zhang88}
and the associated reduction of degrees of freedom is not taken into account.
AL diagrams have been used in the past for calculating the effect of 
fluctuations on the conductivity, \cite{aslamasov68} on Raman scattering
\cite{caprara15,kampf92} and phonon anomalies. \cite{liu16}

  In this paper we study charge instabilities of the one-band $t$-$J$ model 
treating the constraint in the mean-field approximation. We also include
Aslamasov-Larkin diagrams as a natural generalization of the 
ladder approximation. These diagrams produce two kinds of charge instabilities:
nematic or $d$-wave Pomeranchuk instabilities with extremely small wave vectors 
as primary, and CDWs with internal $d$-wave symmetries and much 
larger wave vectors as secondary instabilities.  
Our approach uses Green's and vertex functions on the
real frequency axis which allows to treat both low and high temperatures.
We address, in particular, the following  
points which presently are not well understood: Why is the axial and not the 
diagonal charge instability seen in
experiment? Is there more than one charge instability and what happens
to the mean-field flux instability if the AL diagram is taken into account? 

The paper is organized as follows. After an introduction in section I we 
specify the Hamiltonian in section II and discuss some of its properties 
which are relevant for the following. This section also contains a discussion 
of the diagrams taken into account in our calculation, namely, ladder
and sums of ladder diagrams (details are given in Appendix A) and AL diagrams.
The evaluation of these
diagrams is given in section III. In section IV we investigate
$d$-wave charge instabilities as a function of temperature and doping using
only one basis function with $d$-wave symmetry to represent charge 
fluctuations. 
Since the evaluation of the AL diagrams is somewhat involved we present
details of it in Appendix B. Finally, we consider in Appendix C a complete 
set of four basis functions for charge fluctuations and the related 4x4 
susceptibility matrices.  
It is shown that the leading instability has indeed mainly $d$-wave symmetry.
We also discuss the relationship between different definitions of $d$-wave
charge susceptibilities and the set of employed basis functions.  
 
\section{Hamiltonian and choice of diagrams}

We consider the following effective Hamiltonian for electrons moving on a 
square lattice,

\begin{align}
H =& -\sum_{i,j,\alpha} \delta t_{ij} c^\dagger_{i\alpha}c_{j\alpha} \nonumber \\
&+  \frac{J}{4}\sum_{{\langle i,j \rangle}\atop{\alpha,\beta}} (
c^\dagger_{i\alpha}c_{i\beta} \;\;c^\dagger_{j\beta} c_{j\alpha}
- c^\dagger_{i\alpha}c_{i\alpha} \;\;c^\dagger_{j\beta} c_{j\beta}).
\label{t-J} 
\end{align}

$c_{i\alpha}^\dagger$ and $c_{i\alpha}$ are fermionic creation and
annihilation operators for electrons on site $i$ and spin direction
$\alpha$. $t_{ij}$ denote effective hopping amplitudes for electrons between
the sites $i$ and $j$. The second term in Eq. (\ref{t-J}) represents the 
Heisenberg interaction where $J$ is the Heisenberg constant and $\bra ij \ket$ 
denotes nearest-neighbor sites $i$ and $j$. $\delta$ is the hole doping. 
Without the factor $\delta$ 
in the first term
in Eq. (1) $H$ describes a $t$-$J$ model without any restrictions on double
occupancies, see Ref. [\onlinecite{chowdhury14}]. The conventional $t$-$J$ model is obtained 
if a constraint is added to $H$ which excludes double occupancies of lattice
sites.

The effective Hamiltonian Eq. (1)  captures important features of the 
charge excitation spectrum of the $t$-$J$ model. 
The fermionic operators of the $t$-$J$ model act in the Fock space without
double occupancies; they are Hubbard $X$-operators
and not the usual creation and annihilation operators.
In the large-$N$ limit based on $X$-operators\cite{zeyher96a,greco06}
the $t$-$J$ model becomes, however, equivalent to an
effective Hamiltonian written in terms of usual creation
and annihilation operators (see Eq.(12) in Ref.[\onlinecite{greco06}]).
It contains a kinetic term where the hopping is renormalized 
by $\delta$ neglecting a small contribution proportional to $J$.
A second term represents an effective interaction consisting of  
six separable channels.
Channels $3$-$6$ represent the Heisenberg interaction, channels $1$-$2$
originate from the constraint. Channels $1$-$2$ may be neglected in our case
because we are not interested in s-wave fluctuations and because one can show 
that they couple only weakly to the other channels, Ref.[\onlinecite{cappelluti99}]. 
The effective Hamiltonian Eq.(12) in Ref.[\onlinecite{greco06}] 
becomes then our Hamiltonian Eq. (1). The large-N treatment thus replaces the
hard constraint at $N=2$ by the soft constraint at $N=\infty$.   
Calculating the density correlation function with X-operators in the limit
$N$ to $\infty$ and dropping the channels 1 and 2 yields the same result as 
a calculation using Eq. (1) and a  low-order ladder approximation. 
Many calculations of the spin susceptibility
are based on $H$ of Eq. (1) using the random phase approximation for 
the Heisenberg interaction and a renormalized hopping term, see
Refs.[\onlinecite{brinckmann99,norman01b}].

It is convenient to rewrite $H$ in momentum space.  
The first term in the parenthesis in $H$ becomes,
\begin{equation}
H_J = -\frac{1}{4}\sum_{{{\bf p'}{\bf p''}{\bf k}}\atop{\alpha,\beta}}
J({\bf p'}-{\bf p''})c^\dagger_{{\bf p'}+{\bf k},\alpha} c_{{\bf p'},\alpha}
c^\dagger_{{\bf p''},\beta} c_{{\bf p''}+{\bf k},\beta}
\label{HJ}
\end{equation}
with $J({\bf p})=2J(\cos(p_x) + \cos (p_y))$. 
$J({\bf p'}-{\bf p''})$ can be written as a sum over 4 separable kernels,
\begin{equation}
J({\bf p'}-{\bf p''}) = 4J \sum_{r=1}^4 \gamma_r({\bf p'}) \gamma_r({\bf p''}).
\label{JJ}
\end{equation}
The functions $\gamma_r({\bf p})$ are given by 
$\gamma_1({\bf p}) = (\cos(p_x)-\cos(p_y))/2$,
$\gamma_2({\bf p}) = (\cos(p_x)+\cos(p_y))/2$, 
$\gamma_3({\bf p}) = (\sin(p_x)-\sin(p_y))/2$, 
$\gamma_4({\bf p}) = (\sin(p_x)+\sin(p_y))/2$. 
Using the charge variables
\be
n_{r}({\bf k})  = \sum_{{\bf p},\alpha} 
\gamma_{r}({\bf p})c^\dagger_{{\bf p}+{\bf k},\alpha}
c_{{\bf p},\alpha},
\label{n2}
\ee
and inserting Eq. (\ref{JJ}) into Eq. (\ref{HJ}) yields
\be
H_J = -J \sum_{{\bf k},r} n_r({\bf k}) n^\dagger_r({\bf k}).
\label{HJ2}
\ee
Using similar arguments the last term in $H$ is found to be equal to
$-H_J/2$ neglecting small non-local spin interaction terms. 
Altogether the effective Hamiltonian $H$ becomes, 
\be
H = \sum_{{\bf k},\alpha} \epsilon_{\bf k} c^\dagger_{{\bf k},\alpha}
c_{{\bf k},\alpha} + H'.
\label{H'}
\ee
$\epsilon_{\bf k}$ are one-particle energies,
\be
\epsilon_{\bf k} = 
-2t{\delta}(\cos(k_x)+\cos(k_y)) -4t'\delta \cos(k_x) \cos(k_y),
\label{en}
\ee
$t$ and $t'$ denote
hopping amplitudes between nearest and second nearest neighbors,
respectively. $H'$ is the interaction part of $H$ given by 
\be 
H' =  -\frac{J}{2} \sum_{r,{\bf k}} n_r({\bf k})
n_r^\dagger({\bf k}).
\label{HH}
\ee

The Matsubara Green's function describing charge fluctuations 
reads
\begin{equation}
\Pi_{rs}({\bf k},\tau_1-\tau_2) = - \langle {\cal T}(n_r^\dagger({\bf k},\tau_1) 
n_s({\bf k},\tau_2)) \rangle.
\label{dwave}
\end{equation}
$\tau_1$ and $\tau_2$ are imaginary times, ${\cal T}$ 
the time ordering operator and
$\langle...\rangle$ the expectation value. 
$\Pi_{rs}$ is a 4x4 matrix describing the coupling of charge fluctuations 
with the symmetries $r$ and $s$. 
After a Fourier transform with respect to $\tau_1-\tau_2$,
$\Pi$ can be written as $\Pi(k)$ where $k$ denotes both the momentum 
$\bf k$ and the Matsubara frequency $\rmi\omega_n$, $k=({\bf k},\rmi\omega_n)$. 
Using $H'$ and 
performing a sum over ladders one obtains
\begin{align}
\Pi_{rs}(k) =& \Pi^{(0)}_{rs}(k) -\sum_t\Pi^{(0)}_{rt}(k)J\Pi^{(0)}_{ts}(k) +... \nonumber \\
=& \sum_t \Pi^{(0)}_{rt} (1 + J\Pi^{(0)}(k))^{-1}_{ts}.
\label{rpa}
\end{align}

Eq. (\ref{rpa}) is derived in Appendix A for the simplest case where
$\Pi^{(0)}_{rs}(k)$ is given by the unperturbed charge Green's function, 
i.e., by a simple fermionic bubble. More generally,  $\Pi^{(0)}_{rs}(k)$
denotes the irreducible part of the charge Green's function
which consists of all diagrams of $\Pi_{rs}(k)$ which cannot be written as
a matrix product of the $\Pi_{uv}^{(0)}(k)$ times powers of $J$.  
In Appendix C we show that fluctuations in the
$d$-wave channel $r=1$ are much stronger than in the other three channels
and mix only weakly with them. It is therefore admissible to limit
ourselves in the main part of the paper to $r=s=1$ and to 
drop the index 1 altogether. For a different definition of the charge variables see last 
section of Appendix C. 

\begin{figure}
\centering
\includegraphics[width=6.5cm]{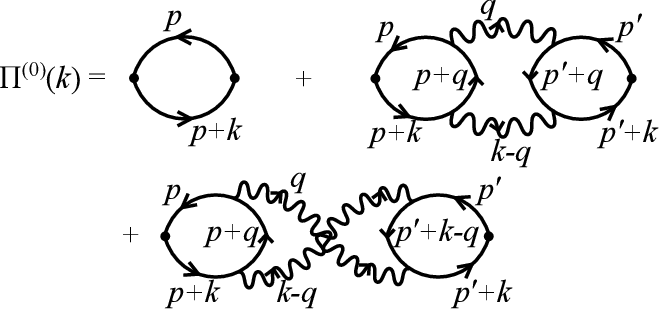}
\caption{
Bubble and Aslamasov-Larkin diagrams for $\Pi^{(0)}$.} 
\label{fig:1} 
\end{figure}

The diagrams which we will take into account for $\Pi^{(0)}(k)$ 
are shown in Fig. \ref{fig:1}. 
Solid lines represent unperturbed electronic Green's functions.   
Small filled circles stand for the d-wave 
vertex $\gamma({\bf p})$, the wavy lines represent spin propagators.   
The physical meaning of the diagrams is rather clear. 
The first diagram  
describes the free propagation of charge excitations between two rungs
of the ladder represented by the small filled circles. We denote it 
by $\Pi^{(0)}_{0}(k)$.
In the second diagram the charge excitation transforms on the way between
two rungs into a product of spin excitations and then back into a charge 
excitation.  
In this way two spin excitations can make an important contribution to 
charge correlation functions.   
Interchanging the two right ends of the spin propagators 
in the second diagram in Fig. \ref{fig:1} 
yields the third diagram where the spin propagators cross
each other. This topologically inequivalent diagram has also to be taken 
into account. We
will call the second and third diagrams Aslamasov-Larkin diagrams and denote 
their sum by $\Pi_{AL}^{(0)}(k)$.    
In the following we are interested in the static limit and put therefore
the external frequency $\rmi\omega_n$ to zero. We also choose the hopping $t$ and
the lattice constant $a$ as energy and length units, respectively, and put
$J = 0.28$ and $t' = -0.3$ throughout the paper.

In the present paper Maki-Thompson processes, as those considered in Ref.[\onlinecite{wang14}], 
are not included. In Ref. [\onlinecite{wang14}] an axial CDW was obtained, however, this
result is controversial\cite{mishra15}. The authors of Ref.[\onlinecite{mishra15}] find that these
processes lead to a diagonal CDW. For this reason and because a proper 
calculation of the Maki-Thompson diagrams are beyond the scope of the 
present paper we focus on AL diagrams for an explanation for the observed 
axial CDW.

\section{Evaluation of the diagrams}

\begin{figure}
\centering
\includegraphics*[angle=270,width=8.5cm]{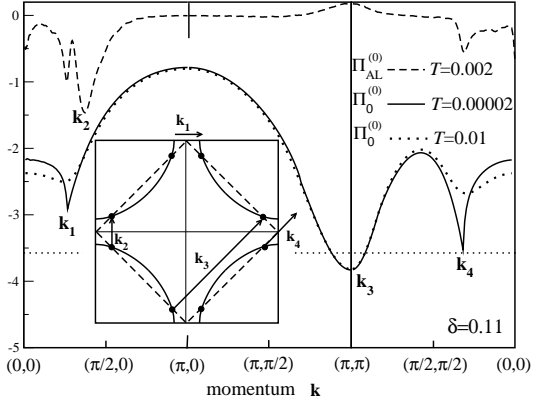}
\caption{(color online)
Momentum dependence of $\Pi_{0}^{(0)}({\bf k})$   
and $\Pi_{AL}^{(0)}({\bf k})$ for doping $\delta = 0.11$. 
${\bf k}_1$ - ${\bf k}_4$ denote nesting vectors or vectors connecting
hot spots, the latter are shown as filled circles in the inset.}
\label{fig:2}
\end{figure}

The first diagram in Fig. \ref{fig:1} is the single bubble contribution 
to $\Pi^{(0)}(k)$ given by, 
\begin{equation}
 \Pi_{0}^{(0)}({\bf k}) = - 2 \sum_{\bf p} \gamma^2({\bf p}) 
\frac{f(\epsilon_{{\bf k}+{\bf p}}-\mu) 
 -f(\epsilon_{\bf p}-\mu)}{-\epsilon_{{\bf k}+{\bf p}} +\epsilon_{\bf p}}.
\label{bubble}
\end{equation}
$f(\epsilon)$ is the fermionic occupation number and $\mu$  the 
chemical potential. 
The solid and dotted lines in Fig. \ref{fig:2} show
$\Pi_{0}^{(0)}({\bf k})$ for $T = 0.00002$ 
and $T = 0.01$, respectively.
The doping is $\delta=0.11$ and the 
momentum $\bf k$ varies from $(0,0)$ to $(\pi,0)$ and $(\pi,\pi)$ and back to 
$(0,0)$. The horizontal, dotted line corresponds to 
$\Pi^{(0)}({\bf k}) = -1/J$ where
the denominator of Eq. (\ref{rpa}) is zero and an instability
of the normal state occurs. The solid line, describing the low-temperature 
case, exhibits two sharp dips at the momenta ${\bf k_{\rm 1}}=(0.832,0)$ and
${\bf k}_{\rm 4}=(0.988,0.988)$. As shown in the inset of Fig. \ref{fig:2} these 
momenta join nested regions along
axial and diagonal directions, respectively. \cite{holder12} Dips due 
to nesting 
are extremely sensitive to temperature and vanish even at moderate 
temperatures as illustrated by the solid and dotted lines in Fig. \ref{fig:2}. 
Figure \ref{fig:2} shows another dip which is rather broad and 
much less sensitive to temperature.
Its wave vector ${\bf k}_{\rm 3} = (\pi,\pi)$ connects two more distant hot 
spots as shown in the inset of Fig. \ref{fig:2}. The Fermi lines near 
these hot spots are not parallel to each other
causing a rather broad and weakly temperature dependent 
$\Pi_{0}^{(0)}({\bf k})$ near ${\bf k}_{\rm 3}$. In spite of the absence
of nesting this instability is rather robust and leads at low
temperatures to the staggered flux state with circulating currents. 
\cite{cappelluti99} 
Figure \ref{fig:2} shows that $\Pi_{0}^{(0)}$ is not able to account for 
the experimental findings in underdoped cuprates: Observed charge 
density waves in axial
direction cannot be ascribed to the low-temperature dip at ${\bf k}_1$
and the predicted flux state in diagonal direction is not seen in the 
experiment. The wave vector ${\bf k}_2$ connecting nearest
neighbor hot spots in vertical or horizontal direction does not 
cause any dip in
$\Pi_{0}^{(0)}({\bf k})$ though it is near to the wave vector of the
experimentally observed charge order. We therefore
propose that higher-order terms to $\Pi^{(0)}({\bf k})$ such as the
AL diagrams should be included.

The calculation of the second and third diagrams, the AL diagrams, 
needs some care in performing the analytic continuation. The usual procedure
to sum over pole contributions leads in general to Green's functions which 
should be evaluated right on the cut. Clearly, the resulting singularities 
must vanish again if all pole contributions are taken into account.
Since it is difficult to achieve such a compensation explicitly
we describe in detail a method in Appendix B where the compensation
is done analytically. 
Our final result for the AL diagrams is,  

\begin{align}
\Pi_{AL}^{(0)}({\bf k}) &= -\frac{3}{8\pi}\sum_{\bf q} 
J^2({\bf q}) J^2({\bf k}-{\bf q}) \int d\epsilon \; 
n(\epsilon) \Big\{ \nonumber\\
& [ \rmIm D({\bf q}, \epsilon) \rmRe D({\bf k}-{\bf q},-\epsilon) \nonumber \\
 &+ \rmRe D({\bf q},\epsilon) \rmIm D({\bf k}-{\bf q},\epsilon) ] \nonumber \\
& \times [ (\rmRe V^S({\bf k},{\bf q},\epsilon))^2  - (\rmIm V^S({\bf k},{\bf q},\epsilon))^2]  \nonumber\\
& +2 \rmIm V^S({\bf k},{\bf q},\epsilon) \rmRe V^S({\bf k},{\bf q},\epsilon) \nonumber \\
 & \times [\rmRe D({\bf q},\epsilon) \rmRe D({\bf k}-{\bf q},\epsilon) \nonumber \\
 &-\rmIm D({\bf q},\epsilon) \rmIm D({\bf k}-{\bf q},\epsilon) ] \Big\}.\;\;\;
 \label{gesamt1}
 \end{align}
$n(\epsilon)$ is equal to $1/(\rmexp(\epsilon/T)-1)$ and
$V^S$ is the symmetrized vertex given by
\begin{align}
 V^S({\bf k};{\bf q},\epsilon) = &V({\bf k};{\bf q},\epsilon) + 
{\tilde V}({-\bf k};{-\bf q},-\epsilon),
 \label{VS1}
\end{align}
with
\begin{equation}
 V(k;q) = T\sum_p \gamma({\bf p})G(p)G(p+k)G(p+q),
 \label{V}
\end{equation}
\be
{\tilde V}(k;q) = T\sum_p \gamma({\bf p})G(p)G(p+k)G(p+k-q).
\label{Vtil}
\ee
An explicit expression for $V(k;q)$ is given in Eq. (\ref{V2}).

Using the symmetrized vertex $V^S$ both AL diagrams 
are taken into account. $D({\bf k},\epsilon)$ is the spin response function
given in Refs. \onlinecite{sherman04,vladimirov09,letacon11}   
\be
D({\bf k},\epsilon) = \frac{A(1-\gamma_{\bf k})}{\omega_{\bf k}^2 -
\epsilon^2 -\rmi\Gamma \epsilon},
\label{D}
\ee
with $\gamma_{\bf k} = (\cos(k_x)+\cos(k_y))/2$, $\omega_{\bf k}^2 = 
4J^2(1-\gamma_{\bf k})(1+\gamma_{\bf k}+1/(4\xi^2))$. $\xi$ is the magnetic
correlation length given approximately by $\xi^2=1.6/\delta$, 
\cite{vladimirov09} $\Gamma$
is a phenomenological damping chosen to be 0.6 and $A$ a constant to be 
determined from the
spin sum rule. Equation (\ref{D}) describes well the experimental spin 
correlation function over
a large doping region and represents a simplified version of the 
theoretically derived expressions 
in Refs. \onlinecite{sherman04,vladimirov09}. For the numerical evaluation
of the momentum sums in Eqs. (\ref{gesamt1}), (\ref{V}) and (\ref{Vtil})
we used 50x50 or 100x100 nets in the Brillouin zone, for the integration
over $\epsilon$ in Eq. (\ref{gesamt1}) about 1200 points.  

The dashed curve in Fig. \ref{fig:2} shows the momentum dependence of
$\Pi_{AL}^{(0)}$ at low temperature $T=0.002$. $\Pi_{AL}^{(0)}({\bf k})$
is over a wide momentum region small compared to $\Pi_{0}^{(0)}({\bf k})$,
especially along the directions $(\pi,0)$ - $(\pi,\pi)$ - $(0,0)$. 
The smallness of $\Pi_{AL}^{(0)}({\bf k})$ along the diagonal 
can be understood in the following way:
The main contribution in the sum over $\bf q$ in Eq. (\ref{gesamt1}) comes 
from the 
region ${\bf q} \approx {\bf Q} = (\pi,\pi)$ and 
$\approx {\bf Q}-{\bf k}$ if $\xi$ is much 
larger than 1. $\bf q$ may then be fixed in 
$V^S({\bf k},{\bf q},\epsilon)$ to these values. Applying a reflection
along the diagonal to the vertices $V$ and $\tilde V$ in Eqs. (\ref{V}) 
and (\ref{Vtil})
reproduces $V$ and $\tilde V$ except for a minus sign due to the form factor
$\gamma({\bf p})$. As a result $V$, $\tilde V$ and thus also $V^S$ 
and $\Pi_{AL}^{(0)}$ are zero if ${\bf k}$ lies along the diagonal.
Allowing for the momentum dependence of the vertices 
$\Pi_{AL}^{(0)}({\bf k})$ is not exactly zero there but still small.
As a result diagonal instabilities are not much enhanced by the AL diagram
in our one-band model. This is different from the three-band model of
Refs. \onlinecite{yamakawa15,tsuchiizu16}  where the AL diagram enhances both 
axial and diagonal charge fluctuations. 
The biggest modification of $\Pi_{0}^{(0)}$ by the AL diagram occurs in the
axial direction between $(0,0)$ and $(\pi/2,0)$. Both $\Pi_{0}^{(0)}$
and $\Pi_{AL}^{(0)}$ show dips near the nesting 
vector ${\bf k}_1$. However, there is an additional dip
in $\Pi_{AL}^{(0)}$ which is present only in $\Pi_{AL}^{(0)}$ and occurs 
at ${\bf k}_{\rmmin}$ near the vector ${\bf k}_2$ (see the inset in 
Fig. \ref{fig:2})
which connects nearest neighbor hot spots in the vertical or horizontal
direction.

\section{Charge instabilities}

\begin{figure}
\centering
\includegraphics[angle=270,width=8.5cm,trim=50 20 30 50,clip]{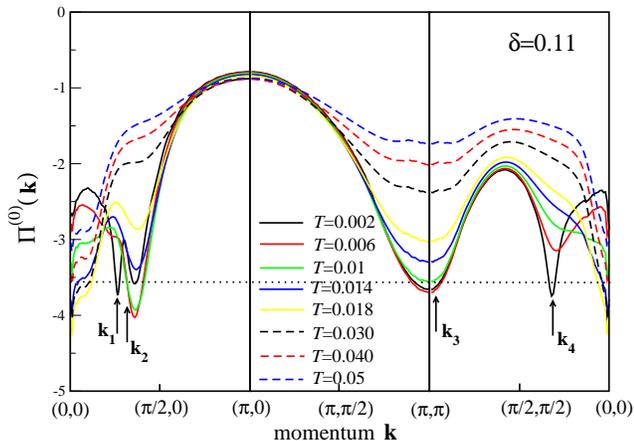}
\caption{(color online)
Momentum dependence of $\Pi^{(0)}({\bf k})$ for different temperatures
and $\delta$ = 0.11.}
\label{fig:3}
\end{figure}

\begin{figure}[h]
\centering
\includegraphics*[angle=270,width=8.5cm,trim=70 20 30 50,clip]{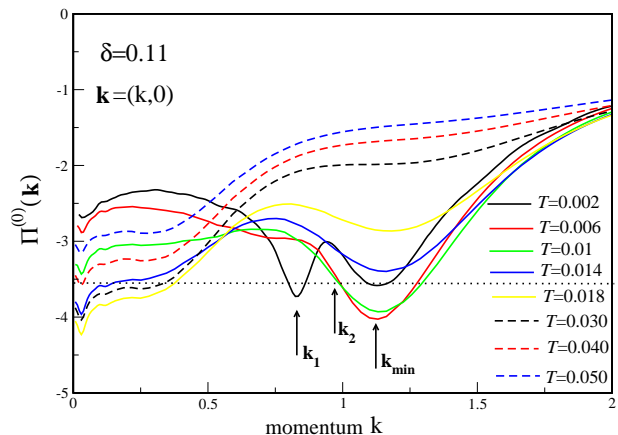}
\caption{(color online)
Enlarged Fig. \ref{fig:3} for axial momenta between $(0,0)$ and $(2,0)$.}
\label{fig:4}
\end{figure}

The total irreducible charge correlation function is given by
\be
\Pi^{(0)}({\bf k}) = \Pi_{0}^{(0)}({\bf k}) + \Pi_{AL}^{(0)}({\bf k}).
\label{sum}
\ee
Figure \ref{fig:3} shows the momentum dependence of $\Pi^{(0)}({\bf k})$
throughout the Brillouin zone for 8 different temperatures. 
Disregarding the axial direction and the region very close to $\Gamma$ 
there are two minima near
the momenta ${\bf k}_{\rm 3}$ and ${\bf k}_{\rm 4}$ which were present already in
$\Pi_{0}^{(0)}$, see Fig. \ref{fig:2}. 
The minimum with momentum ${\bf k}_{\rm 4}$ is due to nesting, 
shifts rapidly upwards with
increasing temperature and soon ceases to be even a local minimum.
The minimum with momentum ${\bf k}_3$ is broad, moves first with increasing 
temperature a little downwards and then slowly upwards. It does not change 
much its shape and remains a local minimum at all temperatures shown. 
Figure \ref{fig:4} shows the same curves as in Fig. \ref{fig:3} but in the 
small interval
between  $(0,0)$ and $(2,0)$. The minimum of $\Pi^{(0)}({\bf k})$
located near ${\bf k}_{\rm 1}$
is extremely sensitive to temperature and exists no longer above 
$T = 0.006$. There is another minimum in axial direction at ${\bf k}_{\rmmin}$,
see Fig. \ref{fig:4}. Its doping dependence and its closeness to 
${\bf k}_{\rm 2}$ (see the inset of Fig. \ref{fig:6})
suggest that it originates from transitions between nearest neighbor 
hot spots. Since in this case the nesting is only approximate the
transitions involve a large region of momentum states around the hot
spots. As a result ${\bf k}_{\rmmin}$ deviates substantially from ${\bf k}_{\rm 2}$
and the temperature dependence around ${\bf k}_{\rmmin}$ is rather weak similar
as in the case of the flux state dip near ${\bf k}_{\rm 3}$. 
$\Pi^{(0)}({\bf k}_{\rmmin})$ lies very near to the normal state at low 
temperatures, decreases first substantially with increasing temperature and
then increases and moves back into the normal state. A similar reentrant 
behavior is found for $\Pi^{(0)}({\bf k})$ at small k near ${\bf k}=(0.03,0)$, 
see Fig. \ref{fig:4}. It exhibits a
minimum which moves first down from the normal state deep into the 
unstable region, reaches there an absolute minimum and then moves upwards 
again into the normal state. Interesting is that the relative or absolute
minimum of $\Pi^{(0)}({\bf k})$ never occurs exactly at ${\bf k} = 0$.
This means that the homogeneous state is unstable against a modulation with
a perhaps small, but finite wave vector ${\bf k}$.

The transition temperatures for charge modulations were obtained by 
starting at high temperatures and then lowering the 
temperature until minima of $\Pi^{(0)}({\bf k})$ cross the dotted line
at $-1/J \sim -3.57$. 
The first crossing occurs at a very small momentum ${\bf k} \approx (0.03,0)$
and a temperature of about $T_{\rmnem}\sim 0.04$ and describes the 
transition to a nematic state. Due to the reentrant
behavior of $\Pi^{(0)}$ there is another crossing at around 
$T \sim 0.011$ from the nematic back to the normal state.
Lowering further the temperature the dip near ${\bf k}_{\rmmin}$ touches the 
dotted line at around $T_{\rmCDW}\sim 0.0128$ as can be seen in 
Fig. \ref{fig:4}. 
Due to
the reentrant behavior there is also in this case a second crossing of 
$\Pi^{(0)}$
at a very low temperature $T_{\rmCDW} \sim 0.001$ to the normal state.  
For the temperature $T = 0.014$,  $\Pi^{(0)}({\bf k})$ lies except for very small
momenta everywhere above the critical line. This means that it is stable
with respect to other CDW states, in particular, the flux state with 
momentum ${\bf k}_{\rm 3}$. The dips near the perfect nesting vectors 
${\bf k_{\rm 1}}$ and ${\bf k}_{\rm 4}$ can lead to 
instabilities only at very low temperatures as can be
seen from Figs. \ref{fig:3} and \ref{fig:4}. At higher temperatures these 
dips no longer exist.
Figs. \ref{fig:3} and \ref{fig:4} show that the CDW state forms in general in the 
presence of the nematic
state. This means that the band stucture of the nematic state and not that
of the normal state should be used in calculating the CDW instabilities.  
Below we will show that the nematic distortion enhances the CDW instabilities
but that in a first approximation this effect may be neglected.

\begin{figure}[h]
\centering
\includegraphics*[angle=270,width=8.5cm,trim=80 20 30 50,clip]{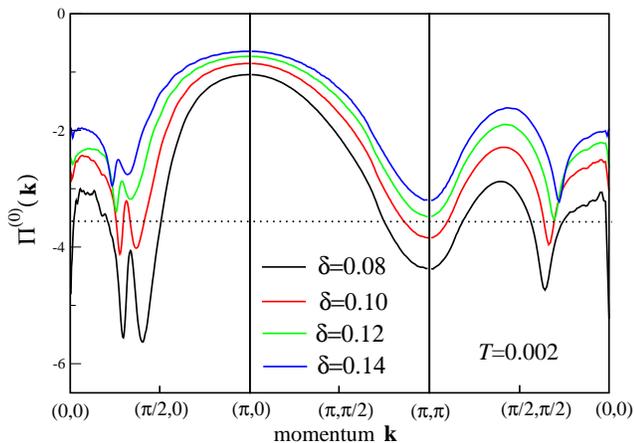} 
\caption{(color online)
Momentum dependence of $\Pi^{(0)}({\bf k})$ at $T=0.002$ for different
dopings.}
\label{fig:5}
\end{figure}

Figure \ref{fig:5} shows the momentum dependence of $\Pi^{(0)}({\bf k})$ at a 
low temperature for different dopings. Interesting is that the axial 
minima outside of the nematic region are much lower than
the diagonal ones for $\delta = 0.08$. With increasing doping this difference
becomes smaller. At around $\delta = 0.12$ the lowest axial and diagonal
dips have about the same depth. Increasing $\delta$ further the diagonal
minima would be lower than the axial ones. However, all the minima would lie
then above the dotted line and thus would not cause any instabilities. 

The transition
temperatures can be determined in the same way as in Fig. \ref{fig:4} for 
$\delta = 0.11$.
The result is shown in Fig. \ref{fig:6} by squares for the nematic and 
by circles for CDW states. Calculated transition temperatures have been
smoothly joined by dashed curves and consist of two branches 
due to the reentrance behavior. Reentrant behavior has also been found
previously in a model with forward scattering and BCS pairing 
interactions \cite{yamase07a} and in the Hubbard model. \cite{pietig99} 
The modulated state exists in each case 
between the
upper and lower branch. If no reentrance occurs the lower branch is 
given by the x-axis. Figure \ref{fig:6} indicates that in our approach 
nematic and CDW states  
are related and direct consequences of the AL diagrams. Compared with 
the experimental phase diagrams in Fig. 18 of Ref. \onlinecite{choiniere18}
and Fig. 3 of Ref. \onlinecite{sato}
and disregarding the reentrant behavior essential features of the 
experiments are roughly reproduced.
For instance, the strong, approximately linear decrease of 
$T_{\rmnem}$ with doping and the approach of $T_{\rmnem}$ to 
the weakly doping dependent $T_{\rmCDW}$ 
from above. The reentrant behavior has so far not been reported experimentally
and is at present a theoretical prediction.
The filled and empty circles in the inset of Fig. \ref{fig:6} 
show the doping dependence of the length of ${\bf k}_2$ 
and of ${\bf k}_{\rmmin}$, respectively. The
squares, triangles and diamonds are experimental data. 
The filled circles lie always a little below the empty circles. This means
that in momentum space an extended region around the hot spots contributes 
to the CDW instability. This also explains the weak temperature
dependence of $\Pi^{(0)}({\bf k})$ around  ${\bf k}_{\rmmin}$, compared to the 
regions around the minima near ${\bf k_{\rm 1}}$ and ${\bf k_{\rm 4}}$.

\begin{figure}[h]
\centering
\includegraphics*[angle=0,width=8.0cm]{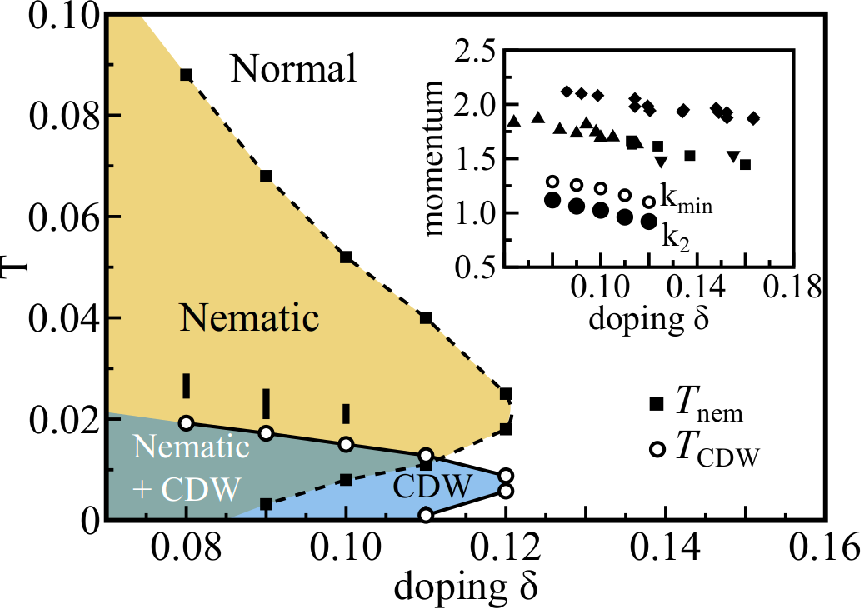}
\caption{(color online)
Nematic and CDW transition temperatures $T_{\rmnem}$ and
$T_{\rmCDW}$, respectively. For the meaning of the vertical bars at the dopings $0.1$, 
$0.09$ and $0.08$ see the text. Inset: Length of ${\bf k}_{\rm 2}$ (filled circles),
${\bf k}_{\rmmin}$ (empty circles) and experimental values for YBCO \cite{tabis17}
(squares), Bi2201 \cite{peng16} (triangles), and Hg1201 \cite{tabis17} 
(diamonds).}
\label{fig:6}
\end{figure}

\begin{figure}[h]
\centering
\includegraphics*[width=7.5cm,angle=-90]{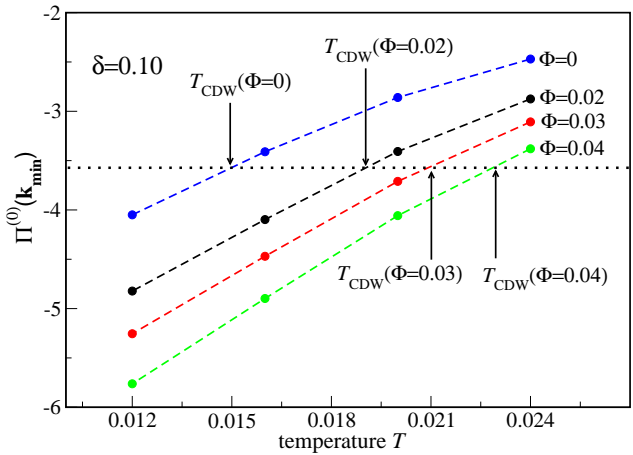}
\caption{(color online)
$\Pi^{(0)}({\bf k}_{\rmmin})$ as a function of $T$ for several nematic distortions
$\Phi$.}  
\label{fig:7}
 \end{figure}

The three empty circles in Fig. \ref{fig:6} at the dopings 0.10, 0.09 
and 0.08 denote
CDW instabilities which lie in the nematic phase. This means that the 
change of the band structure due to the nematic distortion should be
taken into account in calculating these CDW instabilities. Defining
the nematic distortion by 
\begin{equation}
\Phi = -\sum_{{\bf p},\alpha} \gamma({\bf p}) \bra c^\dagger_{{\bf p}\alpha} 
c_{{\bf p}\alpha} \ket,
\label{phi}
\end{equation} 
we have calculated $\Pi^{(0)}({\bf k})$ for a fixed distortion $\Phi$ 
as a function of the axial momentum $\bf k$.  $\Pi^{(0)}({\bf k})$
shows a well-pronounced  minimum 
near the momentum ${\bf k}_{\rmmin}$ defined in Fig. \ref{fig:4}. The temperature 
dependence of this minimum is shown in Fig. \ref{fig:7}. It crosses the 
critical line
(dotted line) at the temperature $T_{\rmCDW}(\Phi)$ which defines the CDW
transition temperature for this $\Phi$. The value for 
$T_{\rmCDW}(0)$ can be read off in Fig. \ref{fig:6} and is also shown in 
Fig. \ref{fig:7}. Fig. \ref{fig:7}
shows that $T_{\rmCDW}(\Phi)$ is a monotonically increasing function of $\Phi$,
at least for the considered interval [0,0.04]. This implies
that a nematic distortion always increases the CDW transition temperature
and thus strengthens the CDW state. This conclusion also agrees with a recent
calculations in the three-band model. \cite{kawaguchi} Fig. \ref{fig:7}
implies that
the largest transition temperature is obtained for the largest possible value 
for $\Phi$. Restrictions for the variable $\Phi$ come from the condition 
that  $\Pi^{(0)}({\bf k})$ lies for all momenta k above 
the critical line. This  restricts $\Phi$ approximately to the interval 
$0.02 < \Phi < 0.04$ . The monotonic increase of $T_{\rmCDW}(\Phi)$ with $\Phi$ 
leads therefore to the inequality 
$$T_{\rmCDW}(0.02) < T_{\rmCDW}(\Phi) <T_{\rmCDW}(0.04)$$ and thus 
to lower and upper bounds for the observable 
transtion temperature $T_{\rmCDW}$. We have inserted these lower and upper bounds
in Fig. \ref{fig:6} for each of the three dopings in form of vertical 
bars connecting these bounds.  
The bars lie above the line which connects the open circles 
which represent $T_{\rmCDW}(\Phi=0)$. They are not far away from the empty
circles for our considered dopings so that the empty circles may be 
considered as a reasonable approximation to the CDW instability also in the
coexistence region.

Two of the filled squares in Fig. \ref{fig:6} lie in the CDW phase
and correspond to instabilities of the CDW state with respect to an
infinitesimally small nematic modulation. To describe these instabilities
correctly one would have first to obtain the self-consistent CDW order parameter
and then to calculate the position where the renormalized 
$\Pi^{(0)}({\bf k} \sim 0) $ crosses the critical line. Since we have not 
determined the CDW order parameter the physical relevance 
of these filled squares remains unclear.

\section{Conclusions}

 In conclusion, we have identified in our approach five different charge
instabilities in the underdoped region. The leading one corresponds to 
a nematic
transition with a transition temperature $T_{\rmnem}$, which decreases 
strongly 
and approximately linearly with increasing doping and vanishes near the doping 
$\delta =0.12$. Below $T_{\rmnem}$ there exist secondary charge 
instabilities with 
wave vectors ${\bf k}_1, {\bf k}_{\rmmin} \approx {\bf k}_2, {\bf k}_3$ 
and ${\bf k}_4$ which are shown in the inset of Fig. 1 and in 
Figs. \ref{fig:3} and \ref{fig:4}. 
${\bf k}_1$ and ${\bf k}_4$ connect nested regions. The associated 
instabilities are extremely sensitive to temperature and do not play a 
role for the phase diagram at finite temperatures. The instabilities
with wave vectors ${\bf k}_{\rmmin}$ and ${\bf k}_3$ are much more robust and
compete with each other. At low doping the minimum in $\Pi^{(0)}({\bf k})$
is much lower near ${\bf k}_{\rmmin}$ than near ${\bf k}_3$. As a result the 
axial CDW with momentum ${\bf k}_{\rmmin}$ will be more stable than the flux phase
with momentum ${\bf k}_3$. With increasing doping the two minima come 
closer and closer until the absolute minimum has moved to ${\bf k}_3$ for 
$\delta$ larger than 0.12. However, both minima are then in the stable region
and neither state represents the ground state. Our calculation leads thus to the
result that the axial instability at ${\bf k}_{\rmmin}$ is always the leading
CDW instability and not the flux or any other diagonal instability. 
This reflects the fact that in our one-band model spin-induced axial
charge fluctuations are substantially larger than diagonal ones. 
Moreover, the leading CDW instability is enhanced 
if it lies in  the nematic phase as shown by the vertical 
bars in Fig. \ref{fig:6}.
Interesting is that in the three-band model of 
Refs. \onlinecite{yamakawa15,tsuchiizu16} there is no flux phase and the 
competition between instabilities 
takes place near the momenta  ${\bf k}_{\rmmin}$ and ${\bf k}_4$
and not ${\bf k}_{\rmmin}$ and ${\bf k}_3$. Other differences are that
the leading instability is in our case a nematic transition with a wave
vector ${\bf k} \sim (0,0)$ and that axial and not diagonal CDWs appear
as leading secondary instabilities without further many-body corrections.

\section{Acknowledgements}

The authors acknowledge useful discussions with M. Bejas, W. Metzner,
H. Yamase, M. Minola, and B. Keimer.
A.G. thanks the Max-Planck Institut f\"ur Festk\"orperforschung
for hospitality. 

\appendix

\section{Ladder summation for $\Pi_{rs}(k)$}

\begin{figure}[h]
\centering
\includegraphics*[angle=0,width=4cm]{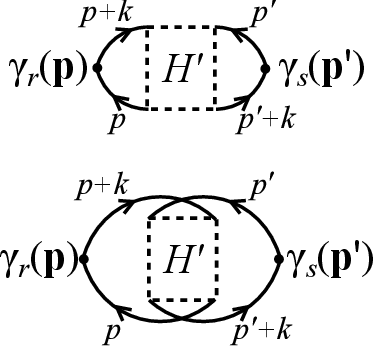} 
\caption{(color online)
First-order ladder diagrams for $\Pi_{rs}({k})$. 
$H'$ is the interaction, 
$\gamma_r({\bf p})$ and $\gamma_s({\bf p})$ are functions defined
below Eq. (\ref{JJ}).}
\label{fig:8}
\end{figure}

Let us consider first the ladder contribution to $\Pi_{rs}(k)$ which is 
linear in $J$.
Using the interaction $H'$ of Eq. (\ref{HH}) two different diagrams can be drawn for
$\Pi_{rs}(k)$ as shown in Fig. \ref{fig:8}. Solid lines denote electron Green's functions,
$\gamma_r({\bf p})$ and $\gamma_s({\bf p})$ are functions defined below 
Eq. (\ref{JJ}). The rectangle originates from the interaction $H'$  and is 
equal to $-J/2\sum_t \gamma_t({\bf p}) \gamma_t({\bf p'})$ for the upper and
equal to $-J/2\sum_t \gamma_t({\bf p'}+{\bf k}) \gamma_t({\bf p}+{\bf k})$
for the lower diagram. Since $\sum_t \gamma_t({\bf p}) \gamma_t({\bf p'})$
depends only on the difference ${\bf p}-{\bf p'}$ the two factors as well as
the two diagrams are equal. This is plausible because interchanging the left 
and right sides of the 
rectangle in the upper diagram produces the lower one and vice versa. 
Adding both diagrams yields 
$-\sum_t \Pi^{(0)}_{rt}(k)J\Pi^{(0)}_{ts}(k)$, i.e., the linear term in $J$ 
in Eq. (\ref{rpa}). $\Pi^{(0)}_{rt}(k)$ is the unperturbed charge Green's 
function. 

Let us consider now the ladder contributions to $\Pi_{rs}(k)$ of order $J^n$.  
There is one diagram of this order where pairs of electrons connect opposite
sides of adjacent rectangles. Interchanging the sides of one or more 
rectangles produces different diagrams which all have the same value 
similar as in the case $n=1$ treated above. Summing over all the diagrams,
the overall prefactor becomes $(-1)^n (J/2)^n\times 2^n = J^n$. Moreover,
the electron lines yield a matrix product of $n+1$ matrices 
$\Pi^{(0)}_{rt}(k)$, i.e.,
we have obtained the $n$-th order term in $J$ of Eq. (\ref{rpa}).

\section{Evaluation of the AL diagrams} 
The second diagram in Fig. \ref{fig:1} 
consists of a product of two identical 
vertices $V(k,q)$ which are connected by
two spin propagators. $V(k,q)$ can be illustrated by a circle 
consisting of three Green's
functions and a bare vertex $\gamma({\bf p})$. Analytically, 
$V(k,q)$ is given by  
\begin{equation}
 V(k;q) = \sum_p \gamma({\bf p})G(p)G(p+k)G(p+q).
 \label{V1}
\end{equation}
The sum over $p=({\bf p},\rmi\Omega_n)$ also includes a sum over 
Matsubara frequencies $\rmi\Omega_n$ which can be carried out analytically.
Writing $q=({\bf q},\rmi\nu_n)$ and considering $z=\rmi\nu_n$ as a general 
complex variable we get
\begin{align} 
 V(k;{\bf q},z) =& \sum_{\bf p}\frac{\gamma({\bf p})}{\epsilon_{{\bf p}+{\bf k}} 
-\rmi\omega_n
 -\epsilon_{\bf p}} \nonumber \\
 & \times \left[ \frac{f(\epsilon_{{\bf p}+{\bf k}})-f({\epsilon_{{\bf p}+{\bf q}}})}
 {z-\rmi\omega_n+\epsilon_{{\bf p}+{\bf k}}-\epsilon_{{\bf p}+{\bf q}}} \right. \nonumber \\
 &+ \left. \frac{f(\epsilon_{{\bf p}+{\bf q}})-f({\epsilon_{{\bf p}}})}
 {z +\epsilon_{{\bf p}}-\epsilon_{{\bf p}+{\bf q}}} \right]. 
\label{V2} 
\end{align}  
Restricting ourselves to the static limit $\rmi\omega_n=0$ $V({\bf k};{\bf q},z)$
has the spectral representation 
\begin{equation}
V({\bf k};{\bf q},z) = \int d\epsilon \frac{{\tilde B}
({\bf k};{\bf q},\epsilon)}{z-\epsilon},
\label{V3}
\end{equation}
with the spectral function

\begin{align}
 \tilde{B}({\bf k};{\bf q},\epsilon) &= 
\sum_{\bf p} \frac{\gamma({\bf p})}{\epsilon_{{\bf p}+{\bf k}}
 -\epsilon_{\bf p}} \nonumber \\
 & \times \left[ (f(\epsilon_{{\bf p}+{\bf k}})-f(\epsilon_{{\bf p}+{\bf q}}))
             \delta(\epsilon+\epsilon_{{\bf p}+{\bf k}}-\epsilon_{{\bf p}+{\bf q}}) \right. \nonumber \\
 & + \left.  (f(\epsilon_{{\bf p}+{\bf q}})-f(\epsilon_{{\bf p}}))
             \delta(\epsilon+\epsilon_{\bf p} -\epsilon_{{\bf p}+{\bf q}}) \right].
 \label{V4}
\end{align}

The third diagram in Fig. \ref{fig:1}  is obtained from the second one by
exchanging the right ends of the spin propagators. This does not affect the 
left vertex but changes the right one to
\be
{\tilde V}(k;q) = \sum_p \gamma({\bf p + \bf k})G(p)G(p+k)G(p+q).
\label{V5}
\ee
Another AL diagram can be generated by exchanging the two ends of the 
spin propagators in the left vertex. This leaves the right vertex 
unchanged whereas the left vertex is now given by
${\tilde V}(k;q)$. Furthermore, another diagram is obtained  
by exchanging the two spin propagators
yielding a diagram where both vertices are given by ${\tilde V}(k;q)$. 
The first and fourth diagram are
topologically equivalent and give identical analytic contributions. The same 
holds for the second and
third diagram. Thus, if the symmetrized vertex $V^S$,
\begin{equation}
 V^S({\bf k};{\bf q},z) = V({\bf k};{\bf q},z) + 
{\tilde V}({-\bf k};{-\bf q},-z),
 \label{VS2}
\end{equation}
is used in the second diagram together with a factor 1/2  
both topologically inequivalent diagrams are taken into account.         
From the above also follows that the symmetrized vertex function $V^S$ 
possesses a spectral representation,
\begin{equation}
 V^S({\bf k};{\bf q},z) = \int d \epsilon 
\frac{B({\bf k};{\bf q},{\epsilon})}{z-\epsilon},
 \label{B}
\end{equation}
with a real spectral function $B({\bf k};{\bf q},\epsilon)$. Similarly, the   
spin propagator $D$ (it is independent of the Cartesian index which 
allows to drop it),
can be written as   
\begin{equation}
D({\bf q},\rmi\omega_n) = \int d\epsilon \frac{A({\bf q},\epsilon)}
{\rmi\omega_n -\epsilon}
 \label{DD}. 
\end{equation}
with the spectral function $A({\bf q},\epsilon)$. 
Using the above spectral representations
the expression for the AL diagrams
becomes
\begin{align}
\Pi^{(0)}_{AL}({\bf k}) = & \frac{3}{8} \sum_{\bf q} J^2({\bf q}) 
J^2({\bf k}-{\bf q}) \int 
d \epsilon d \epsilon' dx dx'  \nonumber \\
& A({\bf q},\epsilon)A({\bf k}-{\bf q},\epsilon')
B(x) B(x') S,
\label{AL1}
\end{align}
\begin{equation}
 S=(-1)\oint dz \frac{n(z)}{2\pi \rmi}\frac{1}{z-\epsilon} \cdot 
\frac{1}{z+\epsilon'} \cdot \frac{1}{z-x}
 \cdot \frac{1}{z-x'},
 \label{AL2}
\end{equation}
$n(z)$ is equal to $1/(\rmexp(\beta z)-1)$ and $\oint$ a contour integral 
circling the real
axis in the mathematically positive sense. The  prefactor (-1) in Eq. 
(\ref{AL2}) arises from
changing the sign in the second factor to the right of the sign $\oint$ in 
Eq. (\ref{AL2}). 
$S$ can also be written as an integral over real $z$,
\begin{align}
 S=&(-1)\int_{-\infty}^\infty dz \frac{n(z)}{2\pi \rmi} \left[
 \frac{1}{z-\rmi\eta -\epsilon} \cdot \frac{1}{z-\rmi\eta +\epsilon'} \right. \nonumber \\
 &\cdot \frac{1}{z-\rmi\eta-x} \cdot \frac{1}{z-\rmi\eta -x'} \nonumber \\
 &- \frac{1}{z+\rmi\eta -\epsilon} \cdot \frac{1}{z+\rmi\eta +\epsilon'} \nonumber \\
 &\left. \cdot \frac{1}{z+\rmi\eta-x} \cdot \frac{1}{z+\rmi\eta -x'} \right],
 \label{AL4}
 \end{align}
where $\eta$ is a positive infinitesimal. 

A straightforward evaluation of the contour integral in Eq. (\ref{AL4}) 
in terms of additive pole contributions
along the real axis encounters some difficulties. For instance, 
if we take the pole contribution from the first
factor in Eq. (\ref{AL2}) we have to insert $z=\epsilon$ in the remaining 
factors under the integral. 
The third factor would lead to $V^S({\bf k};{\bf q},\epsilon)$, i.e., 
a Green's function where the frequency
lies exactly on the cut. Clearly, such a divergence will be canceled
by a zero in the numerator. Such a compensation is, however, difficult
to achieve, both numerically and analytically. We therefore
present below a way where Green's functions never have to
be taken exactly on the real axis and the compensation of singularities is
done automatically.  
Let us introduce the functions
\be
 g_1^\pm = \frac{1}{z \mp \rmi\eta - \epsilon},
\label{g1}
\ee
\be
 g_2^\pm = \frac{1}{z \mp \rmi\eta + \epsilon'},
\label{g2}
\ee
\be
 g_3^\pm = \frac{1}{z \mp \rmi\eta - x},
\label{g3}
\ee
\be
 g_4^\pm = \frac{1}{z \mp \rmi\eta - x'},
\label{g4}
\ee  
and also 
 \begin{equation}
  h_i^\pm = (g_i^+ \pm g_i^-)/2,\;\;\;\; i=1,..,4.
  \label{hh}
 \end{equation}
It follows that 
\begin{equation}
 g_i^+ =h_i^+ + h_i^-,\;\;\; g_i^- = h_i^+ -h_i^-,\;\;\;\; i=1,..,4,
 \label{hh1}
\end{equation}
and thus 
\begin{equation}
h_i^+ = \rmRe g_i^+, \;\;\;h_i^- = \rmi \rmIm g_i^+, \;\;\;\;i=1,...4.
\label{hh2}
\end{equation}

The content of the large parantheses in Eq. (\ref{AL4}) can be written as,
\be
 \{ \} = \Pi_{i=1}^4 g_i^+ - \Pi_{i=1}^4 g_i^- = 
\label{hh3}
\ee
\be
 \Pi_{i=1}^4 (h_i^++h_i^-) - \Pi_{i=1}^4 (h_i^+- h_i^-) = 
\label{hh4}
\ee
\begin{align}
 2(h_1^- h_2^+ + h_1^+ h_2^-)(h_3^+ h_4^+ + h_3^- h_4^-) \nonumber \\
 + 2(h_3^+ h_4^- + h_3^- h_4^+)(h_1^+ h_2^+ + h_1^- h_2^-).
\label{hh5} 
\end{align}
\{ \} contains only odd powers of the imaginary unit $\rmi$. Taking also the 
prefactor $1/(2\pi \rmi)$ into account
it follows that $\Pi^{(0)}_{AL}({\bf k})$ is real, as it should be. The first term in 
Eq. (\ref{hh5}) represents pole contributions from spin propagators, the 
second one from vertices.

To evaluate the expression in Eq. (\ref{hh5})  we split it into 
several contributions:
\begin{align}
(A)&\;\;\;\;\; 2h_1^-h_2^+(h_3^+ h_4^+ + h_3^- h_4^-) = \nonumber \\
&2\rmi \rmIm g_1^+\cdot \rmRe g_2^+ (\rmRe g_3^+\cdot \rmRe g_4^+-\rmIm g_3^+\cdot \rmIm g_4^+).
\label{1}
\end{align}
Taking the complete expression into account we find the following 
contribution to $\Pi^{(0)}_{AL}({\bf k})$: 
\begin{align}
 \frac{3}{8}\sum_{\bf q}& J^2({\bf q}) J^2({\bf k}-{\bf q}) \int 
d\epsilon \; n(\epsilon) A({\bf q}, \epsilon) \rmRe D({\bf k}-{\bf q},-\epsilon) \nonumber \\
 &\times \left[ (\rmRe V^S({\bf k},{\bf q},\epsilon))^2 
- (\rmIm V^S({\bf k},{\bf q},\epsilon))^2 \right],
 \label{A}
\end{align}
where retarded boundary conditions are applied to the Green's 
functions $D$ and $V^S$.

\begin{align}
(B)&\;\;\;\;\; 2h_1^+h_2^-(h_3^+ h_4^+ + h_3^- h_4^-) = \nonumber \\
&2 \rmRe g_1^+\cdot \rmi \rmIm g_2^+ (\rmRe g_3^+\cdot \rmRe g_4^+-\rmIm g_3^+\cdot \rmIm g_4^+).
\label{2}
\end{align}
The resulting contribution to $\Pi^{(0)}_{AL}({\bf k})$ is,
\begin{align}
 \frac{3}{8}\sum_{\bf q}& J^2({\bf q}) J^2({\bf k}-{\bf q}) 
\int d\epsilon \; n(\epsilon) \rmRe D({\bf q}, \epsilon) A({\bf k}-{\bf q},\epsilon) \nonumber \\
&\times \left[ (\rmRe V^S({\bf k},{\bf q},\epsilon))^2 
- (\rmIm V^S({\bf k},{\bf q},\epsilon))^2 \right].
 \label{AA}
\end{align}

\begin{align}
(C)&\;\;\;\;\; 2h_3^+h_4^-(h_1^+ h_2^+ + h_1^- h_2^-) = \nonumber \\
&2 \rmRe g_3^+\cdot \rmi \rmIm g_4^+ (\rmRe g_1^+\cdot \rmRe g_2^+-\rmIm g_1^+\cdot \rmIm g_2^+).
\label{3}
\end{align}
The resulting contribution to $\Pi^{(0)}_{AL}({\bf k})$ is,
\begin{align}
 \frac{3}{8}\sum_{\bf q}& J^2({\bf q}) J^2({\bf k}-{\bf q}) 
\int d\epsilon \; n(\epsilon) B(\epsilon) \rmRe V^S({\bf k},{\bf q},\epsilon) \nonumber \\
 \times& \left[ \rmRe D({\bf q},\epsilon) \rmRe D({\bf k}-{\bf q},\epsilon) \right. \nonumber \\
 &\left. - \rmIm D({\bf q},\epsilon) \rmIm D({\bf k}-{\bf q},\epsilon) \right].
 \label{AAA}
\end{align}

\begin{equation}
(D)\;\;\;\;\; 2h_3^-h_4^+(h_1^+ h_2^+ + h_1^- h_2^-).
\label{4}
\end{equation}
This contribution is identical with that of case (C).
Adding all contributions yields
\begin{align}
\Pi_{AL}^{(0)}({\bf k}) &= \frac{3}{8}\sum_{\bf q} J^2({\bf q}) 
J^2({\bf k}-{\bf q}) \int d\epsilon \; n(\epsilon) \Big\{
\nonumber\\
& \left[A({\bf q}, \epsilon) \rmRe D({\bf k}-{\bf q},-\epsilon) 
+\rmRe D({\bf q},\epsilon) A({\bf k}-{\bf q},\epsilon)\right] \nonumber \\
&\times \left[(\rmRe V^S({\bf k},{\bf q},\epsilon))^2 - (\rmIm V^S({\bf k},{\bf q},\epsilon))^2\right] \nonumber\\
&+ 2B(\epsilon) \rmRe V^S({\bf k},{\bf q},\epsilon) \nonumber \\
 &\times \left[ \rmRe D({\bf q},\epsilon) \rmRe D({\bf k}-{\bf q},\epsilon) \right. \nonumber \\
&- \left. \rmIm D({\bf q},\epsilon) \rmIm D({\bf k}-{\bf q},\epsilon)\right] \Big\}
 \label{gesamt2}
 \end{align}

\begin{figure}[h]
\centering
\includegraphics*[angle=270,width=8.5cm,trim=80 40 30 50,clip]{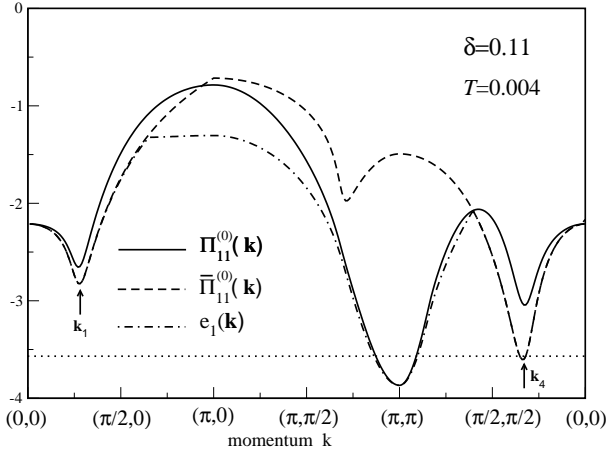} 
\caption{(color online)
Lowest-order (bubble) contribution to $\Pi^{(0)}_{11}({\bf k})$,
$\bar{\Pi}^{(0)}_{11}({\bf k})$ and to the lowest eigenvalue
$e_1({\bf k})$ of $\Pi^{(0)}_{rs}({\bf k})$.} 
\label{fig:9}
\end{figure}

\section{Charge fluctuations with general symmetries}

Keeping all 4 symmetry channels the static susceptibility becomes a 4x4 matrix
$\Pi_{rs}({\bf k})$ which satisfies Eq. (\ref{rpa}), i.e.,
\begin{equation}
\Pi_{rs}({\bf k}) = 
\sum_t \Pi^{(0)}_{rt} (1 + J\Pi^{(0)}({\bf k}))^{-1}_{ts}.
\label{matrix1}
\end{equation}
The matrix $\Pi^{(0)}_{rs}$ can be diagonalized by a unitary transformation $U$,
\begin{equation}
\Pi^{(0){\prime}}_{tt}({\bf k}) = \sum_{rs} U_{tr} \Pi^{(0)}_{rs}({\bf k}) U^\dagger_{st}.
\label{dia}
\end{equation}
Eq. (\ref{matrix1}) becomes then
\begin{equation}
\Pi^\prime_{tt}({\bf k}) = \frac{\Pi^{(0){\prime}}_{tt}({\bf k})}{1+J
{\Pi^{(0){\prime}}_{tt}({\bf k})}}.
\label{matrix2}
\end{equation}
A charge instability with symmetry $v$, wave vector
${\bf k}_c$ and transition temperature $T_{\rm c}$ requires that 
$\Pi^{(0){\prime}}_{tt}({\bf k})$ has a minimum and a positive curvature at 
${\bf k}_c$. Furthermore, the temperature $T_{\rm c}$ must be such that
$\Pi^{(0){\prime}}_{tt}({\bf k}_c) = -1/J$. Alternatively, one can say
that at the transition temperature ${T_{\rm c}}$ $\Pi^{(0){\prime}}_{tt}({\bf k})$ 
touches as a function of ${\bf k}$ the critical line $-1/J$ at ${\bf k}_{\rm c}$
from above.

In section IV we limited ourselves to the basis function $r=1$ and considered
$\Pi^{(0)}_{11}({\bf k})$ and not $\Pi^{{(0)}\prime}_{11}({\bf k})$, which is the 
lowest eigenvalue $e_1({\bf k})$ of the matrix $\Pi^{(0)}_{rs}({\bf k})$.
Keeping only $r=1$ is a good approximation if $\Pi^{(0)}_{11}({\bf k})$
and $\Pi^{{(0)}\prime}_{11}({\bf k})$ are close to each other near the 
instabilities. The first instability $M1$ has the wave vector 
${\bf k}_{\rmnem} = (k_{\rmnem},0)$ and the transition temperature $T_{\rmnem}$ which 
implies 
$\Pi^{(0)}_{11}({\bf k}_{\rmnem}) = -1/J \approx -3.571$. For doping $\delta = 0.08$
we obtain $e_1({\bf k}_{\rmnem}) = -3.522$, for $\delta = 0.11$
$e_1({\bf k}_{\rmnem}) = -3.551$. The approximation to keep only $r=1$
is thus extremely well fulfilled.
The second instability $M2$ has the wave vector ${\bf k}_{\rmmin} = (k_{\rmmin},0)$
and the transition temperature $T_{\rmCDW}$ which again implies  
$\Pi^{(0)}_{11}({\bf k}_{\rmmin}) = -1/J \approx -3.571$. For doping 
$\delta = 0.08$
we obtain  $e_1({\bf k}_{\rmmin}) = -4.097$, for $\delta = 0.11$ 
$e_1({\bf k}_{\rmmin}) = -4.025$. The interaction between different 
density variables is at $M2$ substantially larger than at $M1$.
Nevertheless, the $d$-wave component dominates also at $M2$, as can be seen
from the eigenvector belonging to $e_1({\bf k}_{\rmmin})$: For $\delta = 0.08$ we
have $U_{11} = 0.903, U_{21} = -0.265, U_{31} = U_{41} = -0.239$, 
for $\delta = 0.11$
$U_{11} = 0.922, U_{21} = -0.161, U_{31} = U_{41} = -0.249$.
Thus the critical density fluctuation $\sum_r U_{r1}n_r({\bf k})$ is
always dominated by its d-wave part.

Some authors prefer to define charge variables by
\be
{\bar{n}}_{r}({\bf k})  = \sum_{{\bf p},\alpha} \gamma_r({\bf p})
c^\dagger_{{\bf p}+{\bf k}/2,\alpha}
c_{{\bf p}-{\bf k}/2,\alpha},
\label{n3}
\ee
and Green's functions ${\bar \Pi}^{(0)}_{rs}({\bf k})$ 
by Eq. (\ref{dwave}) with $n_r({\bf k},\tau_1)$ 
and $n^\dagger_s({\bf k},\tau_2)$ replaced by ${\bar n}_r({\bf k},\tau_1)$
and ${\bar n}^\dagger_s({\bf k},\tau_2)$, respectively. 
Shifting $\bf p$ to ${\bf p}+{\bf k}/2$ in the sum over ${\bf p}$ and 
decomposing the trigonometric functions in $\gamma_{r}$ one recognizes that
the two sets of density variables are related to each other
by a unitary transformation.
This means that the corresponding static Green's functions 
$\Pi^{(0)}_{rs}({\bf k})$ and $\bar{\Pi}^{(0)}_{rs}({\bf k})$ 
are also related by
a unitary transformation and have the same eigenvalues. The 
condition for charge instabilities is thus independent of the choice of
density variables. This, however,
is in general not true if one truncates the matrices to scalars. For
instance, if one approximates $\Pi^{(0)}_{rs}({\bf k})$ and 
${\bar \Pi}^{(0)}_{rs}({\bf k})$ by the bubble
contribution, $\Pi^{(0)}_{11}({\bf k})$ and
$\bar{\Pi}^{(0)}_{11}({\bf k})$ are given by the solid and dashed lines in 
Fig. \ref{fig:9},
respectively. In the same approximation the dash-dotted line in that figure 
depicts the lowest 
eigenvalue $e_1({\bf k})$  of the matrix $\Pi^{(0)}_{rs}({\bf k})$. 
The three curves are close to each other at small momenta ${\bf k}$, that is,
near ${\bf k}_1$ and ${\bf k}_4$, as one may expect from the definition
of the density variables. However, for momenta ${\bf k}$ 
near $(\pi,\pi)$ only the solid line is close to the dash-dotted line whereas
the dashed line lies much higher and well above the critical value 
$-1/J \approx -3.571$. This means that the flux instability
is lost if $r=1$ and Eq. (\ref{n3}) are used. In this case 
the choice Eq. (\ref{n2}) is certainly preferable to that of   
Eq. (\ref{n3}). 

\bibliography{al14} 
\end{document}